\documentclass[10pt]{spie}  
\usepackage{multirow}
\usepackage{tcolorbox}
\usepackage{graphicx}
\usepackage{subcaption}
\usepackage{tikz}
\usetikzlibrary{fit}% Package imports
\usepackage{amsmath,amsfonts,amssymb}
\usepackage{graphicx}
\usepackage[colorlinks=true, allcolors=blue]{hyperref}
\usepackage{subcaption} % For subfigures and captions
\usepackage{booktabs} % For better table formatting
\usepackage{xcolor} % For color enhancements in figures/tables
\title{A Novel Quantum Augmented Framework to Improve Microgrid Cybersecurity}

\author[a]{Nitin Jha}
\author[a]{Prateek Paudel}
\author[a]{Abhishek Parakh}
\author[b]{Mahadevan Subramaniam}
\affil[a]{Kennesaw State University, Marietta, GA, USA}
\affil[b]{University of Nebraska Omaha, NE, USA}

\authorinfo{Further author information: Send correspondence to Nitin Jha at 
njha1@students.kennesaw.edu}

\begin{document} 
\maketitle

\begin{abstract}
Small modular nuclear reactors (SMRs) are redefining the energy generation landscape by enabling the deployment of modular, scalable, and pre-built power units that can be used to build distributed autonomous microgrids for critical infrastructure and burgeoning AI factories. Often, these microgrids are linked together to provide a resilient, decentralized power generation infrastructure. Consequently, the cybersecurity of microgrids is of critical importance. In this work, we propose a quantum augmented network framework for resilient microgrids. We integrate the ideas of secure quantum networking, quantum anonymous notification, and quantum random number generation to strengthen the integrity, confidentiality, and privacy of microgrid networks. To substantiate the possible benefits of using quantum augmented microgrids, we simulate a practical high-impact classical attack: a traffic analysis and priority-action spoofing campaign that can (1) deanonymize the anonymous notification for a high-priority action, (2) force excessive key usage, and (3) induce harmful allow/block operations at the control level. We quantify how these attacks affect information leakage, spoof acceptance, key sufficiency, and operational outcomes such as latency, deadline misses, unserved energy, etc. This quantum augmented microgrid (QuAM) framework lets us evaluate trade-offs between privacy, availability, and the operational cost of mitigations (cover traffic, verification delays, and key-rotation policies), further paving the path for the study of more nuanced attacks that arise due to the use of quantum-classical integrated frameworks. 
\end{abstract}


\section{INTRODUCTION}
\label{sec:intro} 

Ensuring sustainability in power generation remains a core challenge, requiring careful trade-offs among economic considerations, operational efficiency, and environmental impact. Recently, research and development efforts have increasingly focused on low-carbon and renewable energy sources \cite{michaelson2021review}. However, different renewable sources of energy, for example, solar and wind, have low power density, intermittency, and a lack of dispatch-ready capability, which adds significant complications and challenges. However, the rapid growth of distributed energy resources and the shift toward decentralized grid architectures have positioned \textit{microgrids} as a critical component of modern power infrastructure. Microgrids integrate diverse energy sources, loads, and storage into local power systems that can operate either alongside the main utility grid or independently in islanded mode \cite{parhizi2015state, hatziargyriou2007microgrids}. This operational flexibility has attracted significant interest from both academia and industry, establishing microgrids as a key element of emerging power systems. Moreover, microgrids can improve grid efficiency by coordinating renewable generation, electric vehicles, energy storage, and other distributed energy resources through robust communication infrastructure\cite{bong2022integration}. 

One application of microgrid is to provide access to electrical services to infrastructure that are either disjoint from the main grid, or have poor quality and less reliability, or these infrastructures are by design supposed to be isolated (like secure military bases or other such infrastructure)\cite{hoffman2025microgrid}. Apart from energy generation and storage, modern microgrids rely heavily on communication networks and digital control systems to coordinate distributed resources, manage load balancing, and maintain system stability. These communication infrastructures enable real-time monitoring, distributed control, and automated decision-making across various components of the microgrid \cite{tang2021quantum}. As a result, microgrids have evolved into complex cyber–physical systems in which power system operations are tightly coupled with communication and information technologies \cite{nejabatkhah2020cyber}. While this integration significantly improves operational efficiency and resilience, it also introduces new security vulnerabilities that can be exploited by cyber adversaries. Attacks targeting communication networks, control signals, or system measurements can disrupt normal operations and potentially lead to severe physical and economic consequences.  

\subsection{Cyber-Physical Security Vulnerabilities in Microgrids}

The severe results of these cyber threats to microgrid infrastructures are well documented. The 2015-16 Ukraine power-grid attacks demonstrated that cyber intrusions targeting industrial control systems can cascade into large-scale blackouts affecting hundreds of thousands of consumers\cite{nejabatkhah2020cyber}. Some earlier studies estimate that such a mass blackout across $15$ U.S. states would affect about $93$ million people and would incur a financial cost in between $\$243$ billion to $\$1$ trillion\cite{bindra2017securing, nejabatkhah2020cyber}. In power electronics-intensive smart microgrids, where inverter-interfaced distributed generators are coordinated through tightly coupled cyber-physical networks, cyber-attacks can have particularly severe effects, as malicious commands propagate directly to physical actuators 
with minimal latency, thus causing massive physical layer disruptions\cite{du2020dynamic}. The cyber-physical model of a smart microgrid generally comprises four layers: (1) the physical power system layer, (2) the sensor and actuator layer, (3) the communication layer, and (4) the management and control layer\cite{irmak2023digital}. False data injection (FDI) is one of the most severe cyber threats to a microgrid infrastructure, as this can be used by an adversary to corrupt the sensor telemetry or state estimation data to cause controllers to make incorrect dispatch and load-shedding decisions. Other threats include Man-in-the-Middle (MITM) attacks on the communication links,  replay attacks exploiting message authentication weaknesses, and coordinated multi-node spoofing campaigns that overwhelm per-source rate-limiting defenses\cite{nejabatkhah2020cyber, irmak2023digital, hoffman2025microgrid}. 

For military and other critical defense installations, the stakes are considered to be even higher. The U.S. Army has identified microgrids as a critical enabler of installation energy resilience, to maintain secure power that should work at all times, regardless of outside disasters, be it either man-made (such as war-like scenarios, terrorist attacks, etc.) or natural (such as earthquakes, tornadoes, tsunamis, etc.). Since these isolated microgrids must operate independently during extended utility outages, it makes them both strategically vital and uniquely vulnerable to sophisticated cyber adversaries who may attempt to disrupt energy supply to critical military operations\cite{hoffman2025microgrid, de2020cybersecurity, jamil2021cybersecurity}. 

\subsection{Toward Quantum-Augmented Security for Microgrids}

As it is being established, all current classical cryptographic schemes, be it factoring large prime numbers (RSA) or solving discrete logarithms (Diffie-Hellman), are vulnerable and easily breakable in the presence of quantum computers. Most of the communications in microgrids depend on these classical cryptographic schemes, and thus, become quite vulnerable in the presence of quantum-adversaries\cite{tang2021quantum}. Quantum key distribution (QKD) comes off naturally as one of the most theoretically researched and experimentally validated countermeasures to a quantum-aware adversary. Recent works have demonstrated QKD integration in networked microgrid testbeds, where QKD systems secure the communications between control centers and local controllers \cite{tang2020programmable, tang2021quantum}. However, QKD alone does not address all the threats facing quantum-secured microgrids, and thus other quantum-communication primitives become essential to this research. Beyond key distribution, quantum cryptography primitives such as Quantum Random Number generator (QRNG), quantum secret sharing, quantum anonymous notification protocols, etc., can offer specific security promises that might help improve the security and resiliency of microgrids\cite{aslam2026survey}. However, a comprehensive research about the integration of some of these quantum primitives in already existing microgrid infrastructure is largely unexplored.

In recent times, people have explored the development of quantum-augmented networks (QuANets) as a practical near-term solution to a quantum-secured network, as a pure large-scale quantum network remains impractical due to several hardware limitations in the current time\cite{jha2024ml, jha2025towards, jha2025quantum}. In this work, we present a simulation study for Quantum-Augmented Microgrid (QuAM), which is conducted through a cyber-physical framework consisting of four integrated layers: (1) physical power system, (2) communication network, (3) quantum security infrastructure, and (4) threat-defense interactions. Unlike previous works where only QKD is used in isolation to increase the security of these networks, QuAM explores the impact of several quantum-secured communications primitives on microgrid resilience metrics, including energy-not-served (EENS), attack detection rates, and key pool dynamics under several adversarial conditions. We evaluate multiple attack scenarios, including false data injection (FDI) combined with node-spoofing and coordinated multi-node campaigns, across diverse network topologies and defense tiers. We also evaluate possible latency overload for the integration of quantum services across multiple topologies and different network sizes to evaluate the scalability of such a quantum-augmented network setup. 

This work is organized as follows: Sec \ref{Sec:methods} goes into the details of the simulator setup, discussing the various layers, attack and defense models, and simulation parameters used for the network-study. Sec \ref{Sec:result} discusses multiple results of the simulation studies, including (1) energy generation profiles to standardize the energy-physics of the simulator, (2) attack and defense studies, and (3) scalability studies with respect to extra overhead latency in the network due to different quantum services. Sec \ref{Sec:conclusion} discusses the conclusions of this study and also discusses several possible extensions of this work in the future, and possibly the practical impact of this study in near-term deployments.

\section{Methodology}
\label{Sec:methods}
\begin{figure}[!htpb]
    \centering
    \includegraphics[width=\textwidth]{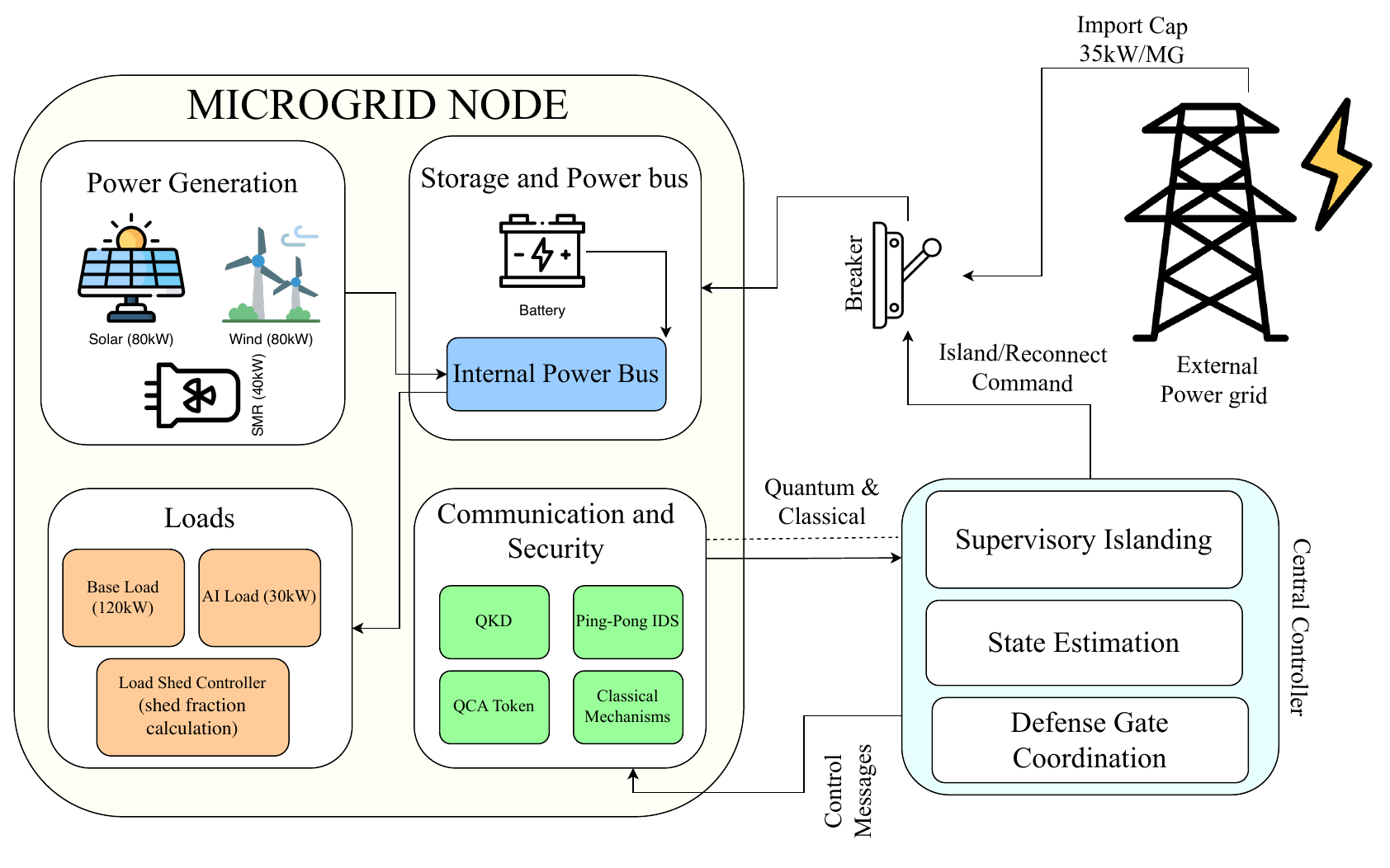}
    \caption{Architecture of a single microgrid unit in the QuAM simulator, showing the four internal subsystems, i.e., generation (SMR, solar, wind), storage and power bus, loads with shed control, and the quantum-secured communications stack (QKD, Ping-Pong IDS, QCA) along with external connections to the central controller via parallel classical and quantum channels, and to the utility grid through a capacity-limited import link ($35$kW per microgrid). For a general simulation, we have five such microgrids under one central controller in different topologies such as star, mesh, ring, and two-cluster bridge.}
    \label{fig:microgriddiag}
\end{figure}

This work evaluates the security and operational resilience of a Quantum-Augmented Microgrid (QuAM) using a discrete-event cyber-physical simulator written in Python 3.11.4. The simulator explores the interaction between four coupled layers:(1) the physical power system, (2) the communication network, (3) the quantum security infrastructure, and (4) the threat-defense decision layer. Together, these layers enable controlled study of how quantum-augmented security mechanisms affect microgrid behavior under normal operation and under adversarial stress. Figure~\ref{fig:microgriddiag} shows a schematic diagram of how a microgrid is designed in our simulator and its interaction with the external power grid through the central controller.

\subsection{Simulator Layers}
As mentioned earlier, QuAM is divided into four layers, which are integrated to form the complete simulator. The layers are as follows,

\begin{enumerate}
    \item[1.]\textbf{Physical Layer:}  models power generation, load, storage, and grid stability.

    Each microgrid node is assigned a heterogeneous energy profile consisting of local generation, demand, and battery storage. Generation is modeled using representative source-specific behavior, i.e., solar output follows a time-varying profile with a configurable peak, wind generation is modeled using a stochastic capacity factor, and small modular reactors (SMRs) provide stable baseload generation. Load is modeled stochastically at each node to capture demand variation over time. Power imbalance is handled through a simplified dispatch and curtailment mechanism. When local generation, storage, and imported power are insufficient to satisfy demand, the controller applies priority-based load shedding. The physical layer tracks system-level and node-level indicators, including cumulative energy not served (EENS), battery state of charge, frequency deviation, and per-node shed fraction. Frequency dynamics are represented using a reduced swing-equation model with configurable inertia and droop parameters so that cyber-induced control failures can propagate into measurable physical instability.

    \item[2.]\textbf{Communication Layer:}
    The communication network interconnects controllers, sensing nodes, and actuators using one of several configurable topologies, including ring, star, mesh, and clustered bridge structures. Each communication edge is parameterized by latency, jitter, bandwidth, and packet-loss probability. Messages are routed over multi-hop paths, and each link includes finite-capacity queueing to capture congestion effects. Three message classes are modeled as follows,
    
    \begin{itemize}
        \item \textbf{PRIORITY ACTION}: emergency shed or restore commands requiring immediate authenticated delivery.
        \item \textbf{CONTROL SETPOINT}: routine dispatch, voltage, or coordination updates.
        \item \textbf{TELEMETRY}: periodic sensor measurements such as generation, load, frequency, and battery state.
    \end{itemize}
    
    These classes differ in urgency, security requirements, and tolerance to delay. As a result, the simulator allows different security protocols and validation pipelines to be assigned according to message criticality.

    \item[3.]\textbf{Quantum Layer:}

    The quantum layer models the availability and consumption of quantum-enabled security resources through an abstracted resource and protocol model. Each eligible link maintains a finite key pool associated with a QKD process, with refill behavior determined by the channel secret-key fraction and current quantum bit error rate (QBER). The simulator updates QBER over time and uses threshold-based logic to represent degraded quantum-channel conditions and possible intrusion alerts, i.e., if QBER increases over a threshold ($11\%$ theoretical bound for ideal BB84 protocol), key-generations are stopped until QBER moves back under this threshold.

    In addition to key generation, the quantum layer supports quantum random number generation (QRNG) for nonce creation, quantum control authentication for message validation, and selected higher-level primitives such as anonymous broadcast and direct secure transmission. These services are not implemented as full physical or protocol-level realizations. Rather, they are represented through mathematical abstractions that capture their functional effects on resource availability, authentication capability, latency overhead, and secure-message handling. Accordingly, the quantum layer is intended to study the system-level implications of quantum augmentation, rather than to emulate every underlying quantum operation at the hardware or circuit level.

    \item[4.]\textbf{Threat-Defense Layer:}

    The threat-defense layer injects attacks at configurable times, durations, and intensities, and processes incoming messages through a multi-stage defense pipeline. The defense model includes access-control filtering, rate limiting, cryptographic verification, token validation, plausibility checks, cross-node consistency checks, and quarantine logic. A message is accepted only if it passes the composite decision policy. This design enables the study of both isolated and coordinated attacks, as well as comparison between no-defense, classical-defense, and quantum-augmented defense configurations.

\end{enumerate}

\subsection{Attack and Defense Models}
The evaluation considers a set of adversarial behaviors designed to span the principal cyber-physical risks faced by networked microgrids:

\begin{itemize}
    \item \textbf{False data injection (FDI)} attacks that manipulate telemetry and distort controller state estimation.
    \item \textbf{Identity spoofing} attacks in which compromised nodes forge privileged control identities.
    \item \textbf{Coordinated multi-node attacks} that distribute malicious traffic to evade local rate limits.
    \item \textbf{Man-in-the-middle (MITM)} attacks targeting message integrity and timing.
    \item \textbf{Key-exhaustion attacks} that attempt to drain QKD-derived key pools through excessive authenticated traffic.
    \item \textbf{Quantum-channel disturbance attacks} that elevate QBER and reduce secure key generation.
\end{itemize}

Attack intensity is varied parametrically to evaluate the robustness of each defense configuration under increasing adversarial pressure. There are also three defense systems in our simulator as follows,

\begin{itemize}
    \item \textbf{No defense}: no authentication, no rate limiting, and minimal verification delay. Base case to study the behavior of the simulator without any defense.
    \item \textbf{Classical defense}: layered filtering, rate limiting, and conventional cryptographic authentication.
    \item \textbf{Quantum-augmented defense}: classical defense mechanisms supplemented with QKD-derived authentication, quantum control validation, channel monitoring, intrusion detection using Ping-pong protocol\cite{bostrom2002deterministic}, and selected quantum communication primitives, such as Kak's three-stage protocol\cite{jha2024joint} for secure direct communication for highest priority messages.
\end{itemize}

\subsection{Topologies Used}
\begin{figure}[ht!]
    \centering

    \begin{subfigure}[t]{0.48\textwidth}
        \centering
        \includegraphics[width=\textwidth]{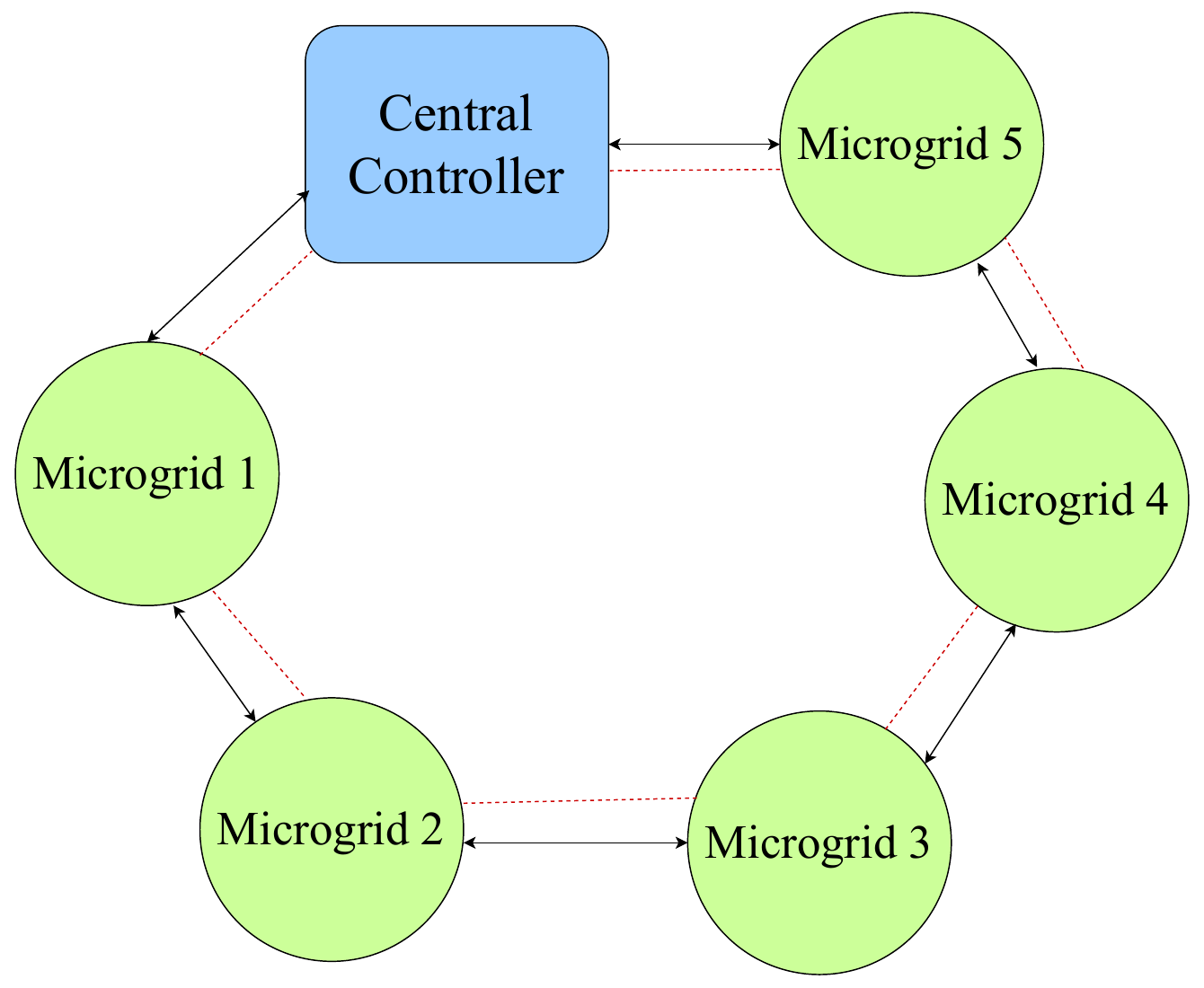}
        \caption{Ring topology}
        \label{fig:ring_topology}
    \end{subfigure}
    \hfill
    \begin{subfigure}[t]{0.48\textwidth}
        \centering
        \includegraphics[width=\textwidth]{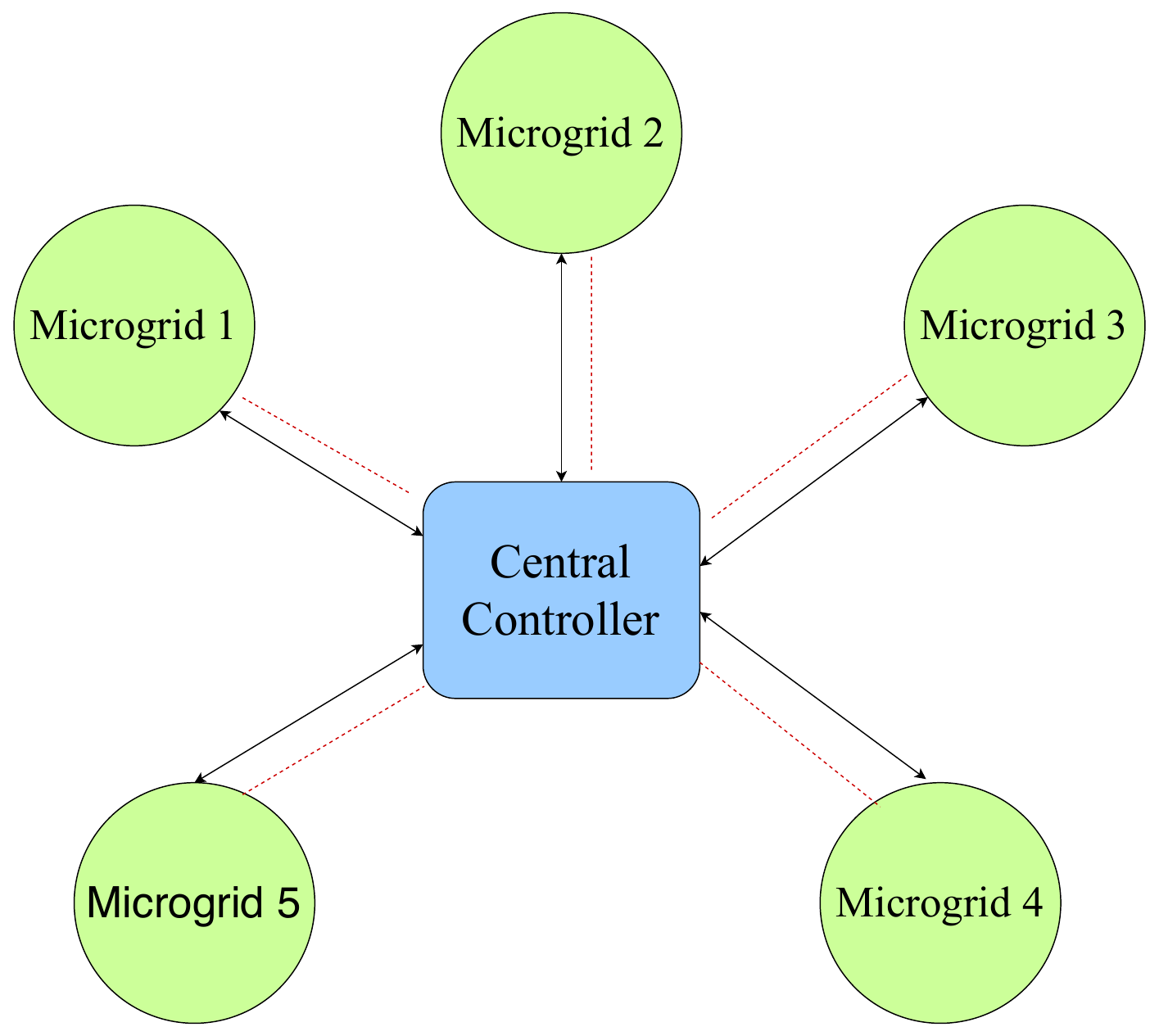}
        \caption{Star topology}
        \label{fig:star_topology}
    \end{subfigure}

    \vspace{0.5em}

    \begin{subfigure}[t]{0.48\textwidth}
        \centering
        \includegraphics[width=\textwidth]{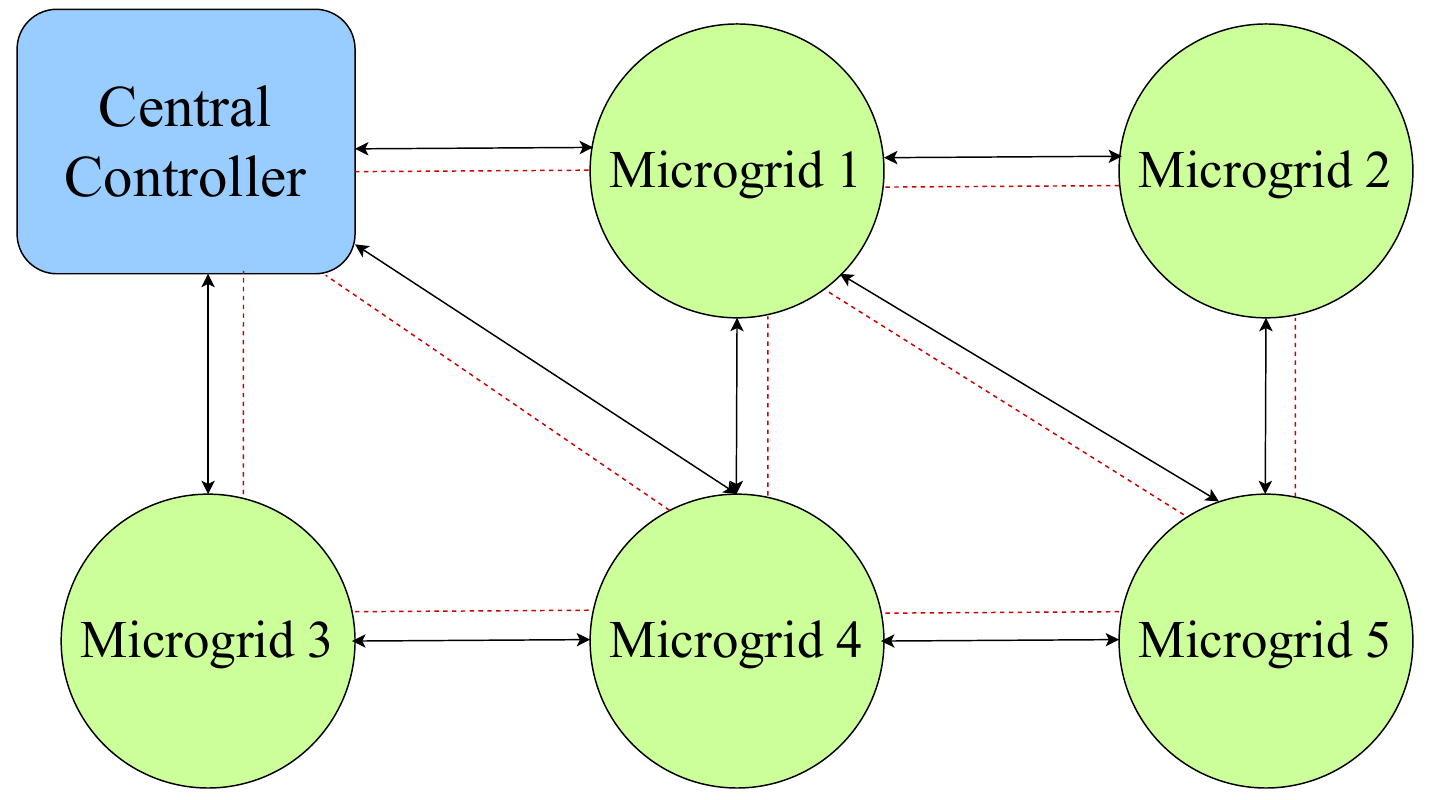}
        \caption{Mesh topology}
        \label{fig:mesh_topology}
    \end{subfigure}
    \hfill
    \begin{subfigure}[t]{0.48\textwidth}
        \centering
        \includegraphics[width=\textwidth]{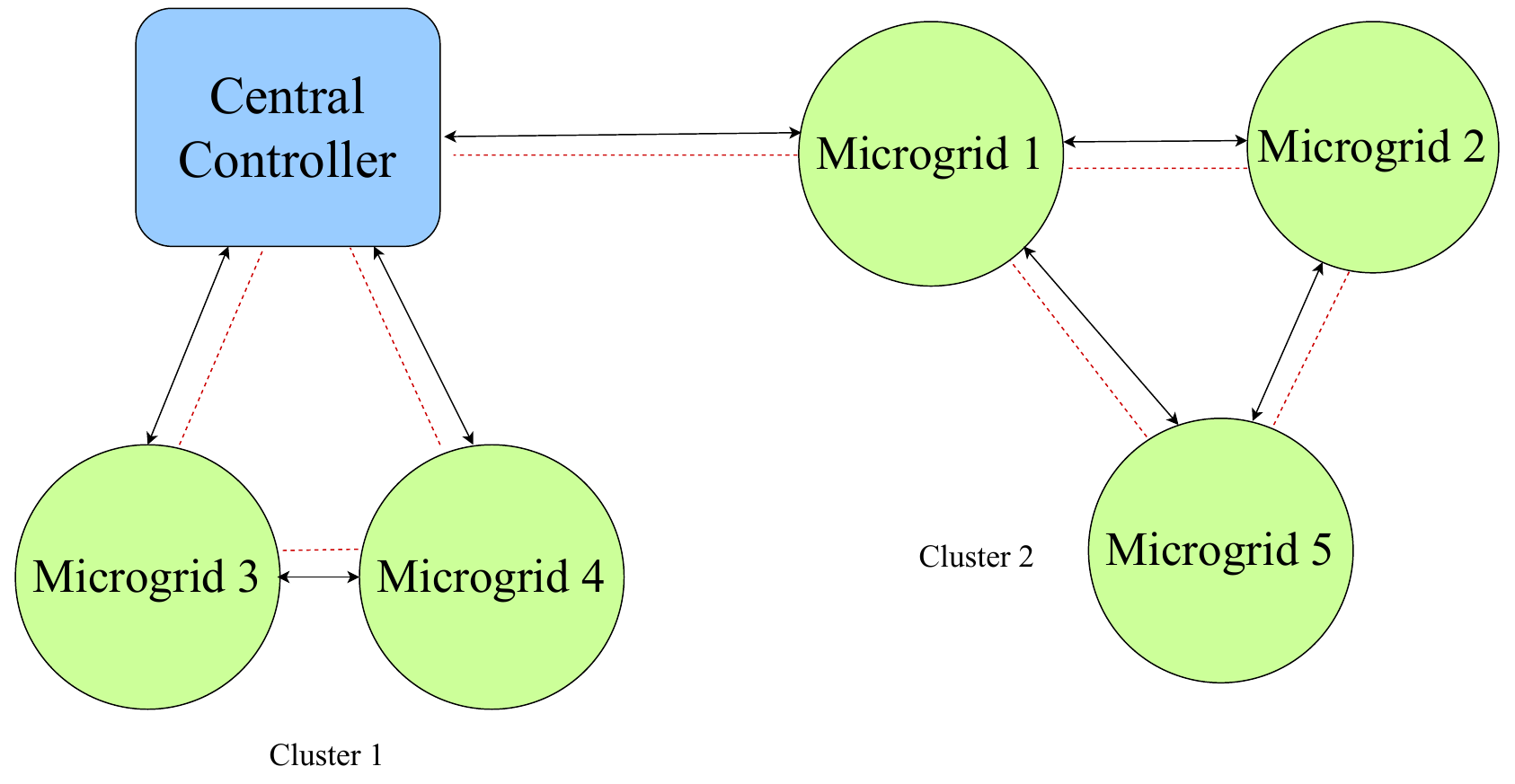}
        \caption{Two-cluster topology}
        \label{fig:tree_topology}
    \end{subfigure}
    \vspace{0.2cm}
    \caption{Network topologies used in the simulator: (a) ring, (b) star, (c) mesh, and (d) two-cluster. These topologies were used to evaluate how structural differences affect communication paths, resilience, and resource usage in the proposed framework. The central controller is responsible to be connect the microgrid network to an external power grid, and more over maintain the proper functionality of each microgrid.}
    \label{fig:topologies}
\end{figure}

For our simulator, we defined four topologies to evaluate different network configurations and communication structures. Figure~\ref{fig:topologies} shows the different network topologies used in our simulator. The central controller is responsible for coordinating the energy-dynamics of the system by sending appropriate commands to maintain appropriate energy levels and distribute energy accordingly. The central controller is also responsible to go into island mode, i.e., if the microgrids are connected to the external grid or stay in islanded mode.

In the star topology, every microgrid node connects directly to the central controller, yielding single-hop control latency but concentrating risk at the hub. The ring topology arranges all nodes in a cycle, providing one redundant path between any pair at the cost of increased network diameter. The mesh topology augments the ring with additional chord links, enabling equal-cost multi-path (ECMP) routing and maximum resilience to simultaneous link failures. The two-cluster bridge topology partitions the network into two fully connected sub-graphs joined by a single inter-cluster link, modeling geographically separated deployments where the bridge becomes a critical bottleneck under attack. Each physical communication edge carries two parallel channels: a classical link for SCADA command and telemetry transport, and a quantum channel supporting a dedicated QKD key pool, authenticated sessions via E91 or three-stage communication sessions, and periodic Ping-Pong IDS probes. Consequently, topology choice governs not only routing latency and fault tolerance but also aggregate quantum key consumption, intrusion detection coverage, and the attack surface available to adversaries targeting key exhaustion or channel eavesdropping.

\subsection{Experimental Design}

The simulator is evaluated across multiple studies, which includes baseline security comparison, scalability analysis, ablation studies, latency trade-off analysis, etc. Across these studies, the design varies in topology, network size, defense configuration, attack scenario, and random seed. Ablation experiments progressively enable defense components to isolate their marginal contribution. Scalability experiments vary the number of nodes and communication topology to study whether security benefits persist as the system grows. This also includes the variation of overhead latency due to added quantum services across an increasing number of nodes. In the next section, we will explore several simulation results.

\section{Results}
\label{Sec:result}
In this section, we will look at some of the results from our simulation study from QuAM. We organize the results into three subsections. First, we validate the physical fidelity of the simulator by examining energy dispatch dynamics, quantum channel behavior, and the theoretical operating bounds of the implemented quantum primitives. Second, we evaluate the security posture across attack types, defense tiers, and network topologies through ablation and escalation analyses. Third, we quantify the operational overhead introduced by quantum security services, including protocol usage scaling, key consumption, and overall latency overhead due to integrated quantum services. All single topology curves are done over Star topology with 5 nodes ($1$ controller and $4$ microgrid nodes) over $60$ mins simulation window with a distributed attack window.

\subsection{Physical Functionality}
\begin{figure}[!htpb]
    \centering
    \includegraphics[width=0.8\textwidth]{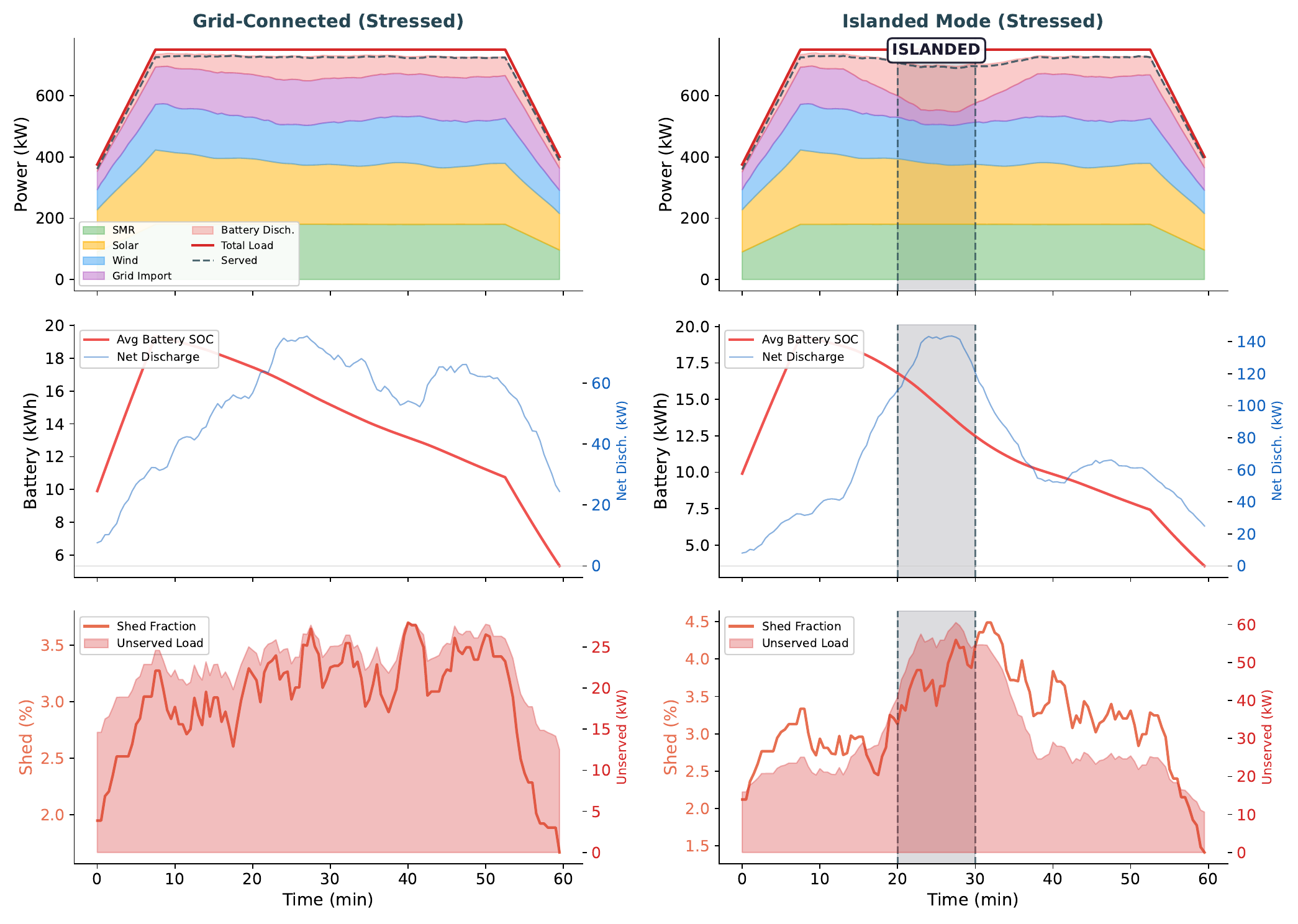}
    \caption{General energy dynamics of the simulator. Two modes shown: grid-connected and islanded mode, i.e., disconnected from the power grid. }
    \label{fig:energydynamics}
\end{figure}

In this subsection, we first present some of the physical workings of the simulator. Figure~\ref{fig:energydynamics} compares microgrid energy dynamics under grid-connected and islanded operation with stressed infrastructure (import cap $35$kW/MG, battery $60$kWh). Even grid-connected, the constrained import capacity produces mild load shedding ($2-3\%$) during periods of low renewable output. During a $10$-minute islanding event ($20$-$30$ min), grid import drops to zero, forcing the battery to compensate, i.e., SOC falls from $15$ to $5$kWh, net discharge spikes to $140$kW, and peak unserved load doubles from $25$ to $60$kW. This establishes the no-attack physical baseline against which cyberattack-induced disconnections are quantified.

\begin{figure}[!htpb]
    \centering
    \includegraphics[width=0.8\textwidth]{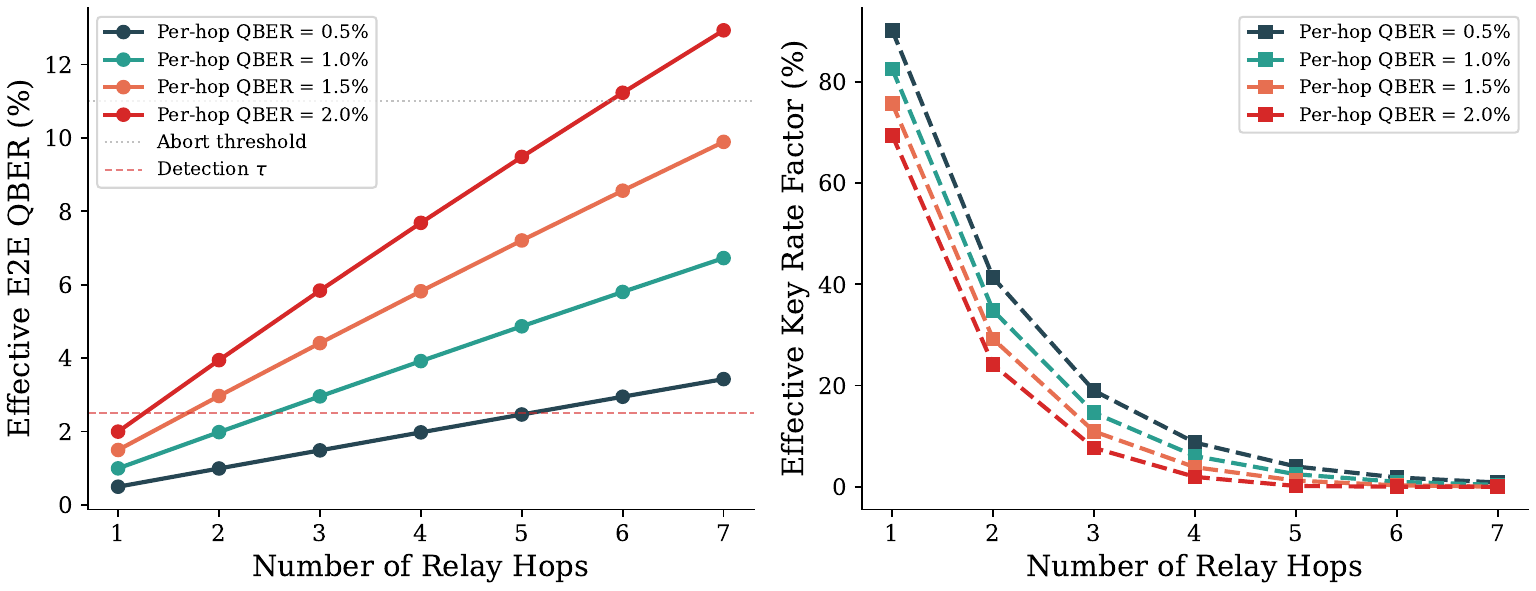}
    \caption{Entanglement swapping scalability: end-to-end QBER accumulation (left) and effective key rate factor (right) as a function of relay hops for four per-hop QBER levels.}
    \label{fig:entanglement}
\end{figure}

Figure~\ref{fig:entanglement} characterizes the scalability limits of entanglement swapping across multi-hop quantum relay paths. The left panel shows that effective end-to-end QBER grows linearly with hop count; at a per-hop QBER of $2\%$, the abort threshold ($11\%$) is reached within 6 hops, while the intrusion-detection threshold ($\tau = 2.5\%$) is breached after a single hop. The right panel shows the corresponding key rate factor, which decays exponentially, dropping below $10\%$ beyond 3-4 hops for all per-hop error rates. Together, these results justify the simulator's $\leq$ 2-hop topology diameter, where entanglement-based authentication remains practical.

\begin{figure}[!htpb]
    \centering
    \includegraphics[width=0.8\textwidth]{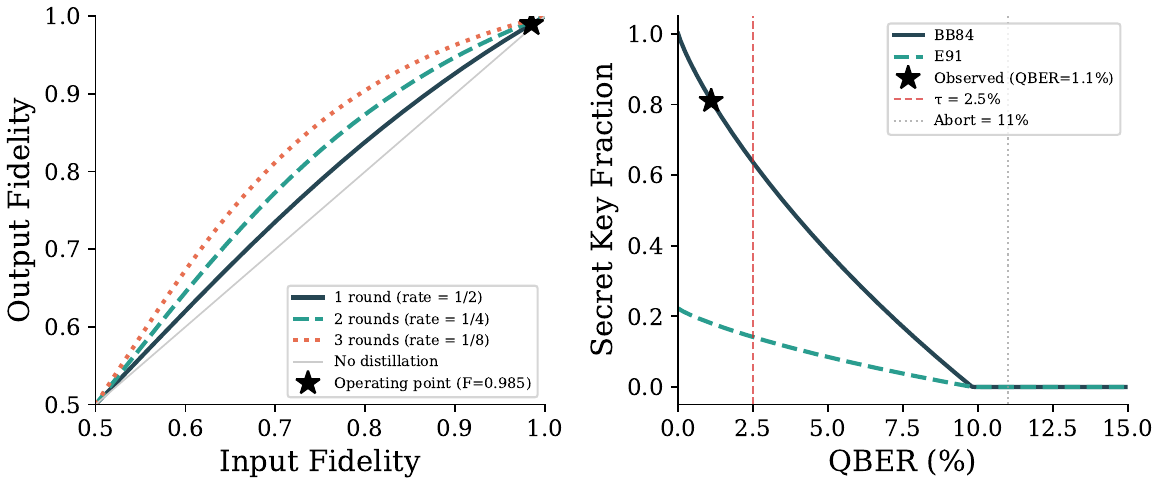}
    \caption{Bennett–Brassard–Popescu–Schumacher–Smolin–Wootters (BBPSSW) entanglement distillation fidelity gain (left) and secret key fraction vs.\ QBER for BB84 and E91 protocols (right).}
    \label{fig:fidelity}
\end{figure}

Figure~\ref{fig:fidelity} presents the theoretical foundations underlying the simulator's quantum key distribution layer. The left panel shows BBPSSW entanglement distillation \cite{bennett1996purification}, i.e., each successive round raises output fidelity at the cost of halving the pair yield, with the simulator's operating point at $F=0.985$ after one round. The right side plots the extractable secret key fraction as a function of QBER for BB84 and E91. At the simulator's observed QBER of $1.1\%$, BB84 retains ${\sim}80\%$ secret key fraction, well below the intrusion-detection threshold $\tau=2.5\%$ and the abort limit of $11\%$. Beyond $\tau$, the Ping-Pong IDS flags anomalous error rates. Beyond the abort threshold, key generation ceases entirely, validating the security margins adopted in the simulation.

\subsection{Attacks and Defenses Evaluation}
Now, we start by inserting attacks into our simulator and studying the energy dynamics. We start by first looking at the energy time series and the effect on the quantum link. 
\begin{figure}[!htpb]
    \centering
    \includegraphics[width=0.9\textwidth]{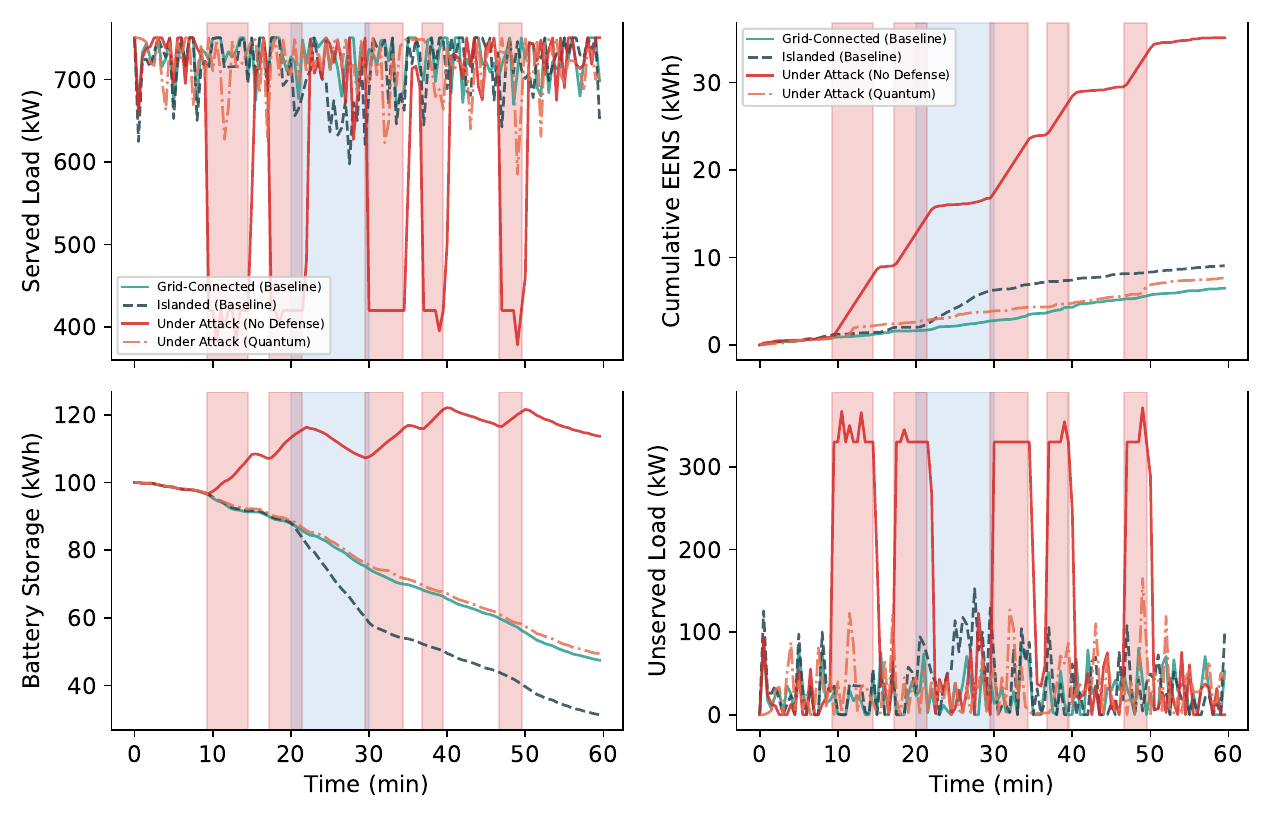}
    \caption{ Energy time-series under four operating conditions: grid-connected baseline, islanded baseline (blue band around $20–30$min), coordinated FDI and node-spoofing attack without defense, and the same attack with quantum defense. Red bands indicate the five S3-intensity attack windows.}
    \label{fig:timeseries}
\end{figure}

Figure~\ref{fig:timeseries} presents energy time-series for four operating scenarios over a 60-minute simulation horizon. The attack scenario employs a coordinated false data injection (FDI) combined with a node-spoofing campaign, in which the adversary simultaneously injects falsified sensor readings (e.g., inflated generation values, manipulated state-of-charge reports) and forges control messages under spoofed node identities. Attacks are launched at intensity level S3, which is the highest tier in the simulator's three-level threat model (S1: reconnaissance/probing, S2: moderate single-vector injection, S3: full-rate multi-vector exploitation targeting both cyber and physical layers simultaneously). Five distributed attack windows (shaded red bands) are spread across the horizon. We see the effect of islanding and grid-connected mode, and the effect of quantum defense classes (QKD + Ping-Pong IDS and QCA authentication) as well. With quantum defense, the attack's impact is substantially mitigated, i.e., served load, and EENS remain close to the islanded baseline, confirming that the quantum security stack effectively contains coordinated cyber-physical threats.

\begin{figure}[!htpb]
    \centering
    \includegraphics[width=0.9\textwidth]{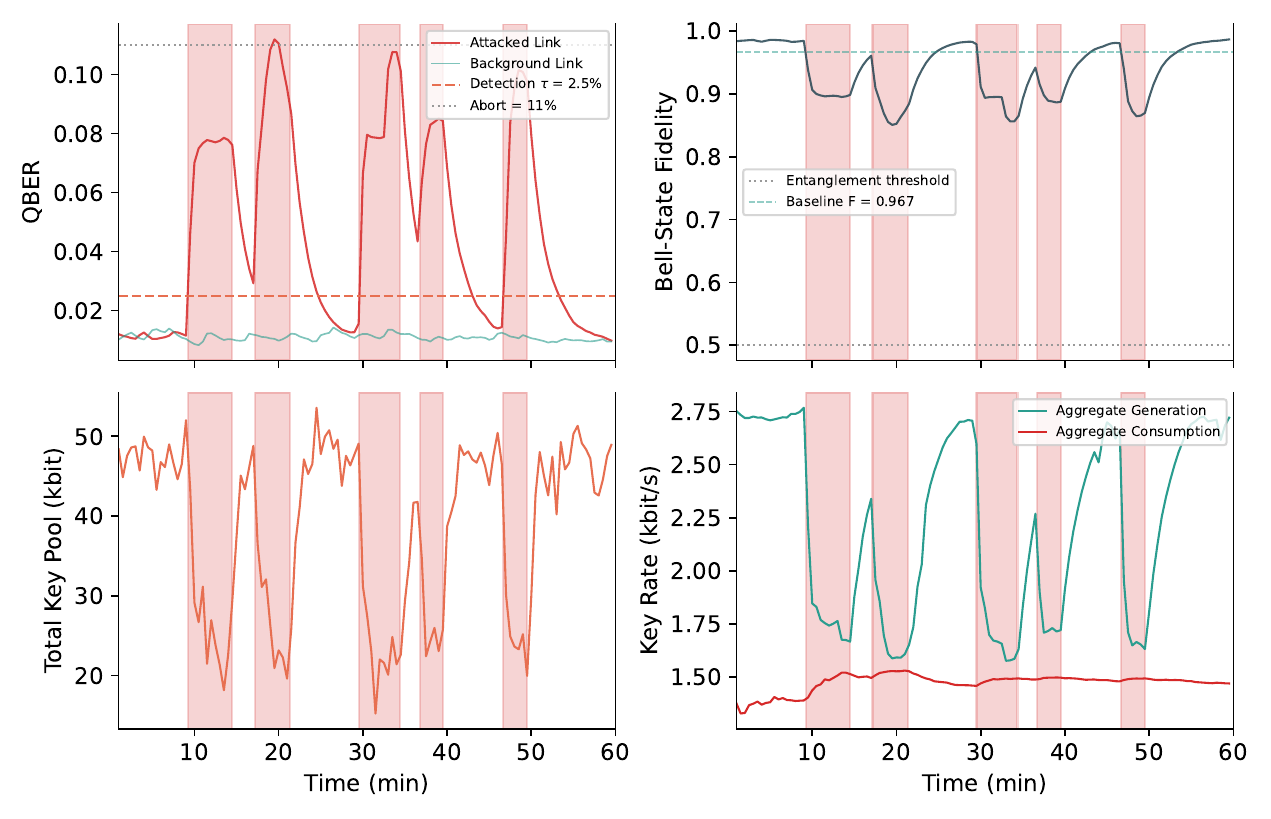}
    \caption{Quantum-channel behavior of an attacked link under quantum defense during the coordinated FDI and node-spoofing attack at intensity S3, showing QBER spikes, fidelity degradation, and temporary QKD key-pool depletion with partial recovery between attack windows (red-shaded areas).}
    \label{fig:qber}
\end{figure}

Figure~\ref{fig:qber} explores the quantum-channel response of a single attacked link under quantum defense during the coordinated FDI and node-spoofing campaign at attack intensity S3. The targeted link’s QBER rises to $8-10\%$ during the five attack windows (shaded red), far above the Ping-Pong IDS threshold of $\tau = 2.5\%$, while a background link remains stable near $1.1\%$. At the same time, Bell-state fidelity drops from $0.967$ to about $0.85-0.90$, but stays above the $0.5$ entanglement threshold, so the channel remains usable. Across the network, the aggregate QKD key pool falls from roughly $50$kbit to $20$kbit during attacks because elevated QBER reduces the secret-key fraction and suppresses key generation, while consumption remains relatively steady. Between attack windows, generation recovers, surplus key production replenishes the pool, and the system demonstrates partial self-recovery under intermittent attack.

\begin{figure}[!htpb]
    \centering
    \includegraphics[width=0.9\textwidth]{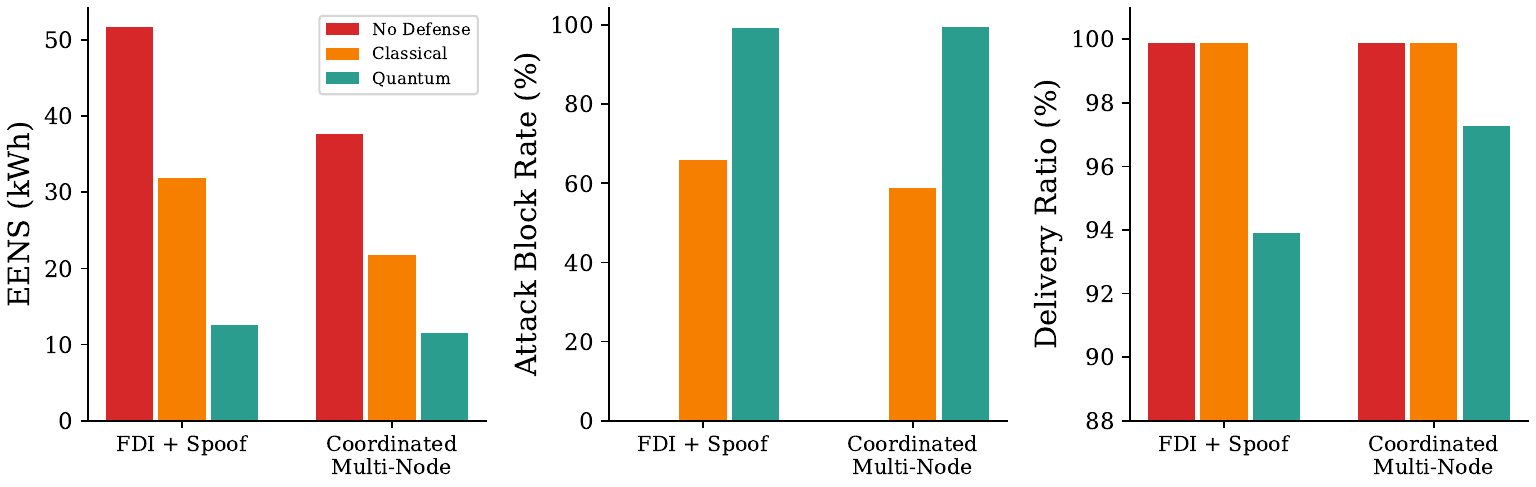}
    \caption{Comparison of the effects of several attacks vs respective defensive over critical quantities such as EENS, attack block rate, and message delivery ratio.}
    \label{fig:attack_type}
\end{figure}

Figure~\ref{fig:attack_type} compares the impact of two attack classes, FDI + node-spoofing and coordinated multi-node, across the three defense tiers. Without defense, FDI + spoofing produces the highest EENS ($52$kWh), while coordinated multi-node attacks yield $38$kWh. Classical defense reduces EENS by $37$\% and $42$\% respectively, achieving block rates of $65\%$ and $60\%$. Quantum defense eliminates nearly all residual impact, reducing EENS to $\sim12$kWh for both attack types and achieving $100\% $ attack block rate. However, the quantum tier's aggressive verification introduces a delivery ratio trade-off: message delivery drops to $94\%$ (FDI and Spoof) and $97\%$ (coordinated), compared to $\sim$$100$\% for no defense, a consequence of the Ping-Pong IDS and QCA token rejection of unverifiable messages.

\begin{figure}[!htpb]
    \centering
    \includegraphics[width=0.9\textwidth]{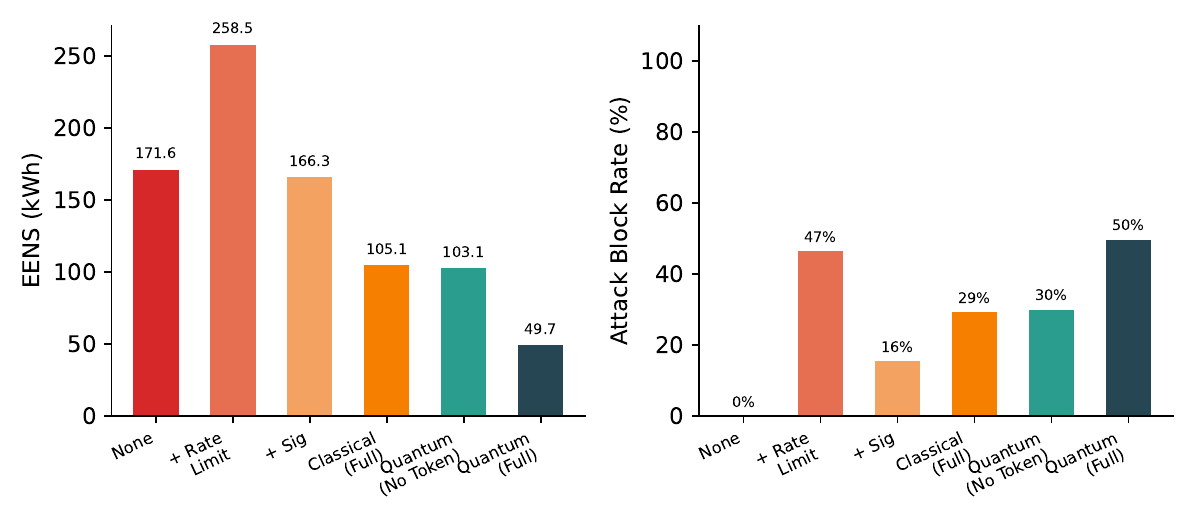}
    \caption{Comparison of the effects of several attacks vs respective defense over critical quantities such as EENS, and attack block rate.}
    \label{fig:abalation}
\end{figure}

Figure~\ref{fig:abalation} presents a defense ablation study, showing the effect of each defense mechanism individually in the presence of all the network attacks.  Starting from no defense (EENS = $172$kWh, $0\%$ block rate), adding rate limiting alone paradoxically increases EENS to $259$kWh, and throttling legitimate control traffic degrades grid coordination more than it impedes the attacker. Restoring signature verification recovers performance ($166$kWh) by filtering forged messages. The full classical stack (rate limiting + signatures + classical IDS) reduces EENS to $105$kWh with a $29\%$ block rate. Upgrading to quantum without QCA tokens yields marginal improvement ($103$kWh, $30\%$), indicating that QKD encryption alone adds limited value when control-plane authentication is absent. The complete quantum stack, adding QCA token authentication, halves EENS to $50$kWh and achieves a $50\%$ block rate, demonstrating that token-based command authentication is the decisive quantum component. Rate limiting increases EENS beyond the no-defense case because, under flooding conditions, it suppresses not only malicious traffic but also legitimate control commands. This unintended throttling degrades coordination and worsens microgrid energy distribution.

\begin{figure}[!htpb]
    \centering
    \includegraphics[width=0.9\textwidth]{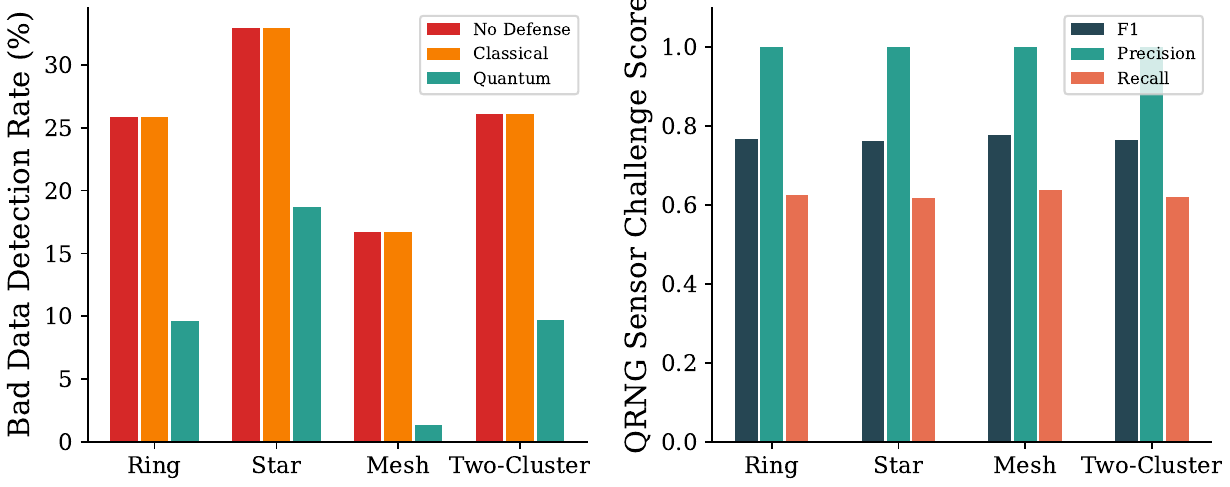}
    \caption{Detection mechanism performance across four topologies: chi-squared bad data detection rate from the WLS state estimator (left) and QRNG sensor challenge classification scores (right).}
    \label{fig:detection}
\end{figure}

Figure~\ref{fig:detection} compares two different intrusion detection mechanisms across all of the four network topologies. The left panel shows the bad-data detection rate of the WLS state estimator\cite{meriem2016study}, which uses a $\chi^2$ test on measurement residuals to flag corrupted readings. Detection is highest in the star topology ($\sim33\%$) because its centralized structure improves observability, and lowest in the mesh topology ($\sim17 \%$) where distributed paths dilute residual sensitivity. Under quantum defense, detection drops near zero across all topologies because falsified telemetry is blocked before reaching the estimator. The right panel shows the QRNG-based sensor challenge mechanism, where the controller issues unpredictable measurement checks and compares reported values against a physics-based model using an adaptive threshold with EWMA (exponentially weighted moving average) tracking. This method achieves perfect precision and about $0.63$ recall, meaning it avoids false accusations while correctly identifying around $63\%$ of compromised sensors. Its main quantum advantage is that QRNG-timed challenges are fundamentally unpredictable to an attacker.

\subsection{Overhead Analysis}
In the previous subsection, we showed that using quantum communication primitives as a defense layer can help us with blocking several classical attacks. However, these defense benefits are not without extra overhead that these quantum primitives can add to the network. In this section, we study some of the overhead costs associated with using quantum primitives as defense layers over classical defense infrastructure for increasing network size across all four network topologies.

\begin{figure}[!htpb]
    \centering
    \includegraphics[width=\textwidth]{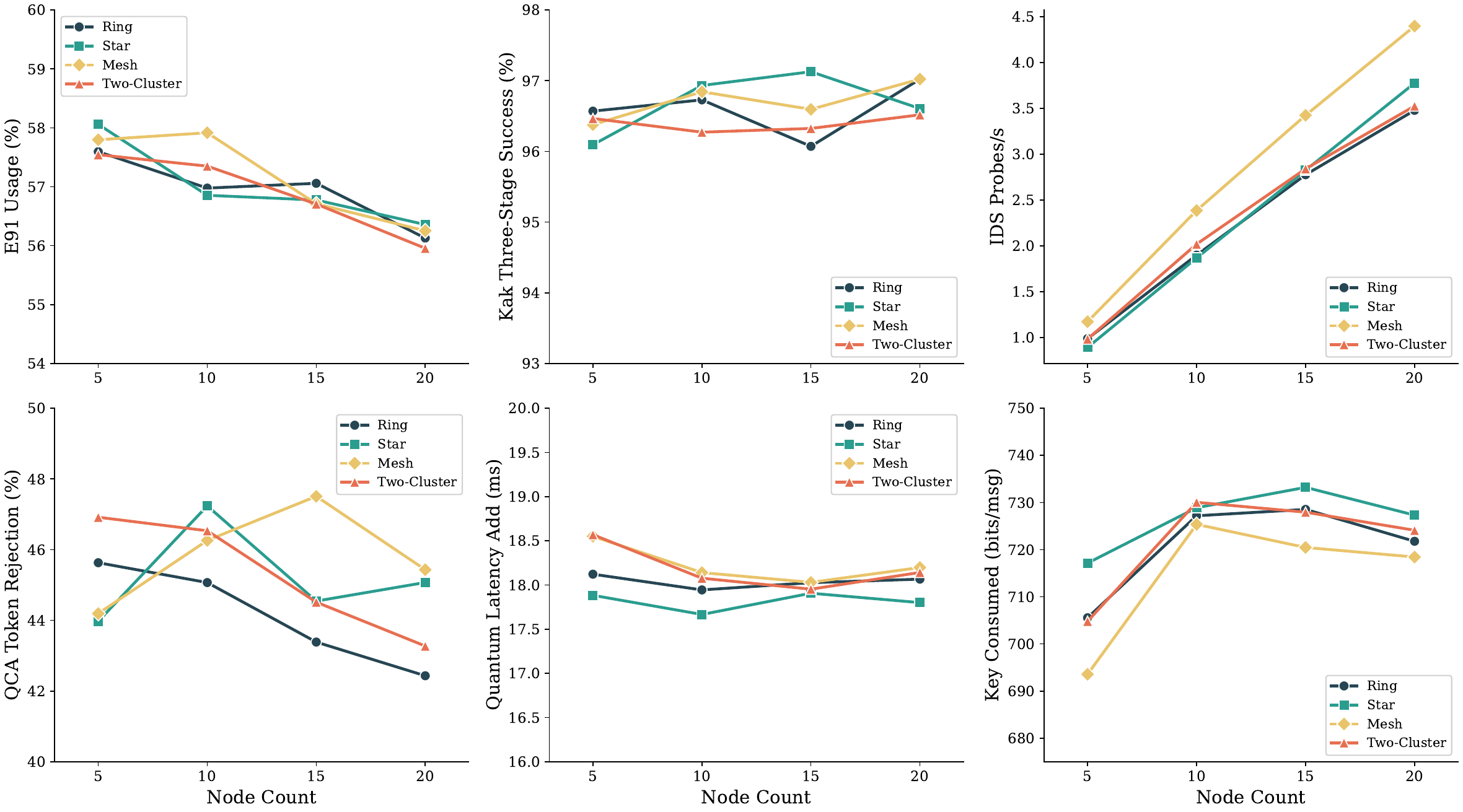}
    \caption{Quantum security overhead as a function of node count ($[5,20]$) across four network topologies. Top row, left to right: E91 entanglement-based key-exchange utilization (\%), Kak's three-stage quantum secure direct communication success rate (\%), and Ping-Pong IDS probe rate (probes/s). Bottom row: QCA token rejection rate (\%), additional latency introduced by quantum verification (ms), and QKD key material consumed per authenticated message (bits/msg).}
    \label{fig:overhead-quant}
\end{figure}

Figure~\ref{fig:overhead-quant} shows six different quantum-communication metrics plotted against the increasing number of microgrids. E91 key-exchange utilization remains stable at $56-58\%$ across all configurations, indicating that the QKD key pool is not exhausted even at $20$ nodes. The Kak three-stage protocol maintains a success rate above $96\%$ at every scale, while the additional quantum verification latency plateaus near $18$ms. IDS probe rate is the only metric that scales approximately linearly, increasing from $\sim1.0$ to $\sim4.0$ probes/s at $20$ nodes as the Ping-Pong protocol must challenge a larger number of peers. QCA token rejection remains between $43-47\%$, showing that a consistent fraction of spoofed control messages is blocked by quantum certificate authentication. Key consumption grows only slightly, from $\sim700 $ to $\sim730$bits/msg, due to additional entanglement-verification rounds rather than message size. Across all six panels, the variation across topologies remains below $3\%$, reinforcing that, at this node scale, security performance is governed more by the defense layer than by the communication topology.

\begin{figure}[!htpb]
    \centering
    \includegraphics[width=\textwidth]{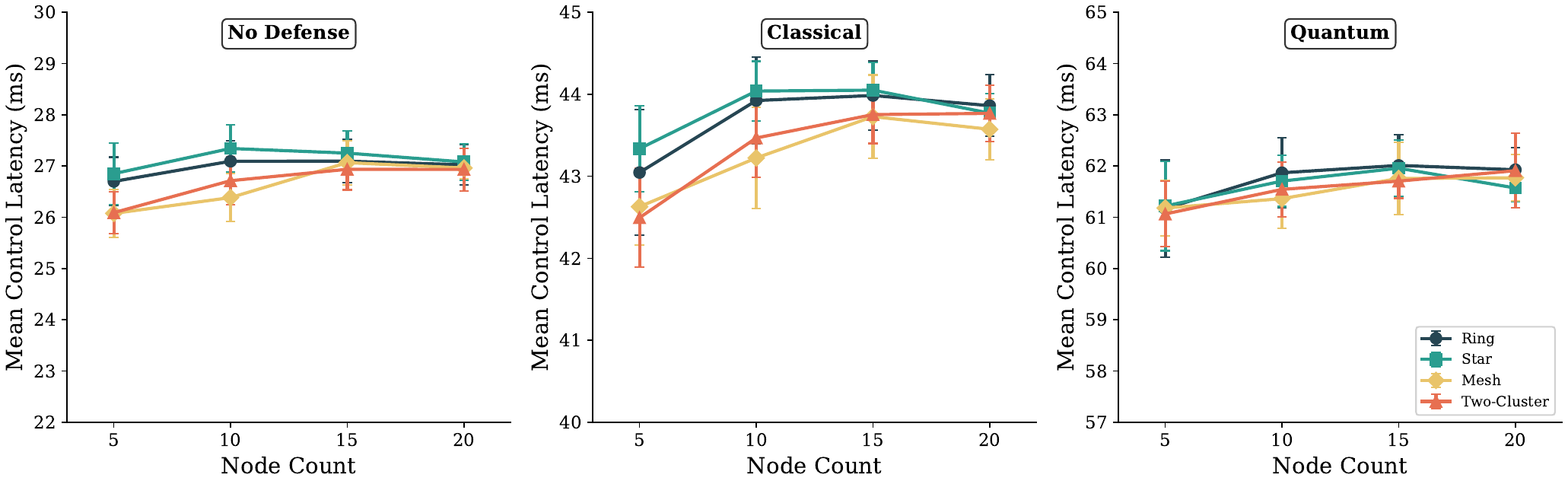}
    \caption{Mean control-loop latency versus microgrid node count (5–20) under three defense tiers across four communication topologies. Left: no defense ($\sim26–28$ms), center: classical authentication ($\sim42–44$ms), and right: full quantum stack comprising QKD, Ping-Pong IDS, and QCA token verification ($\sim61–62$ms).}
    \label{fig:overhead-network-latency}
\end{figure}

Figure~\ref{fig:overhead-network-latency} shows the latency overhead of each defense layer. Without defense, mean latency remains around $26$-$28\mathrm{ms}$ across all topologies. Classical authentication increases this to $42$-$44\mathrm{ms}$, adding a nearly constant overhead of about $16\mathrm{ms}$. The full quantum stack raises latency further to $61-62\mathrm{ms}$, mainly due to QKD key lookup, QCA token validation, and Ping-Pong IDS challenge-response operations. The latency gap between topologies also shrinks with stronger defense: the no-defense case shows up to $2\mathrm{ms}$ difference, while the quantum case keeps all topologies within about $1\mathrm{ms}$, indicating that quantum handshake delay dominates over topology-dependent routing effects.

\begin{figure}[!htpb]
    \centering
    \includegraphics[width=\textwidth]{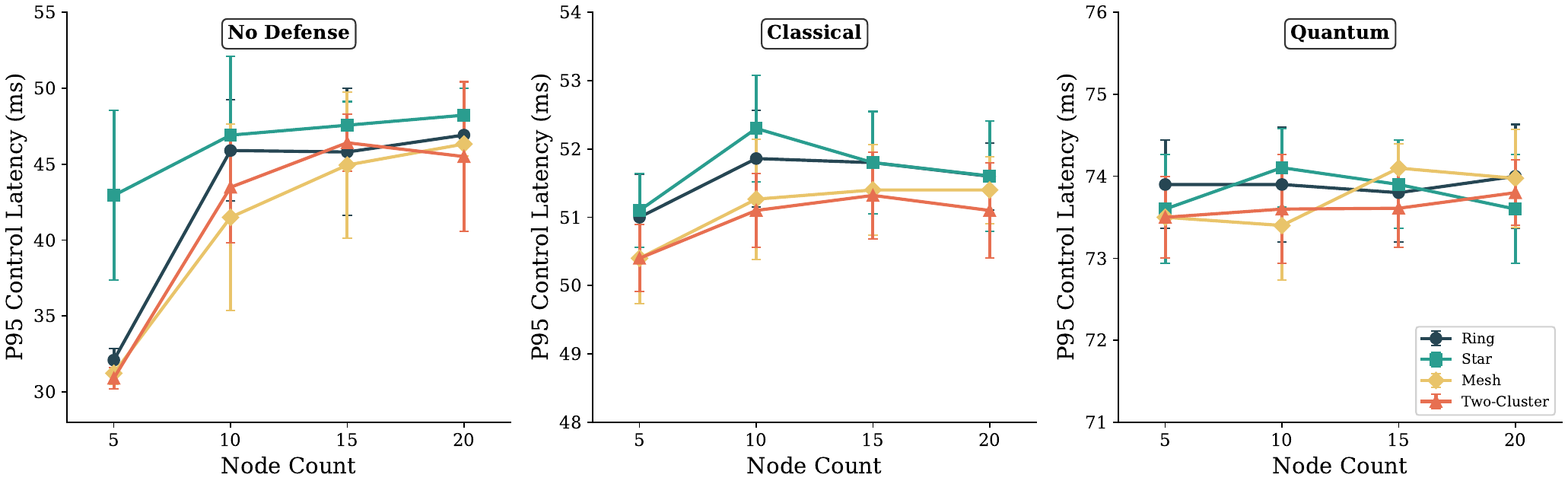}
    \caption{95th-percentile control-loop latency versus node count (5–20) under three defense tiers across four communication topologies. Left: no defense, where tail latency grows from $\sim31$ms to $\sim48$ms with increasing node count; center: classical authentication ($\sim50–52$ms); right: full quantum stack ($\sim73–74$ms). Inter-topology spread compresses from $\sim12$ms (no defense) to $<1$ms (quantum), indicating that quantum handshake overhead dominates and masks topology-dependent jitter. Error bars denote variance over multiple seeds.}
    \label{fig:overhead-latency}
\end{figure}

Figure~\ref{fig:overhead-latency} shows the tail-latency behavior across defense layers. We use P95 (95th percentile) latency, which means that $95\%$ of all control messages were delivered within this time or faster, and only $5\%$ took longer. It captures the worst-case behavior rather than the average. Without defense, P95 latency increases from $31\mathrm{ms}$ at $5$ nodes to $48\,\mathrm{ms}$ at $20$ nodes, indicating greater contention as the network scales. The classical tier limits this growth, with P95 rising only from $50\mathrm{ms}$ to $52\,\mathrm{ms}$. The quantum tier is the most stable, remaining tightly clustered at $73-74\mathrm{ms}$ with less than $1\mathrm{ms}$ variation across topologies, as the QKD and QCA verification pipeline adds a nearly fixed delay. Even at $20$ nodes, quantum P95 stays below $75\mathrm{ms}$, remaining well within the $100$ms control-loop budget for protection-class messaging under stressed network conditions \cite{leggett2010station}.

In summary, the QuAM simulator reproduces physically consistent microgrid behavior in both grid-connected and islanded modes, supporting its use as a cyber-physical testbed. The results show that the full quantum defense stack largely removes the residual attack impact left by classical defenses, with QCA token-based command authentication emerging as the key component in the ablation study. This comes with modest overhead: mean control-loop latency increases by about $35$ms, while $95$th-percentile latency remains below $75$ms across all tested topologies and scales, staying within the $100$ms protection-class messaging budget~\cite{leggett2010station}. Topology has little effect on security or overhead at this scale, suggesting deployment can be guided mainly by physical constraints.

\section{Conclusions}
\label{Sec:conclusion}
In this work, we presented a quantum-augmented microgrid (QuAM) simulator to study the security and operational resilience of microgrids under realistic adversarial conditions. By integrating the physical power layer, communication network, quantum security infrastructure, and threat-defense logic into a single simulator, QuAM enables system-level evaluation of how quantum-enabled security services affect both cybersecurity and microgrid operation. The results show that the simulator reproduces physically consistent behavior under both grid-connected and islanded operation, providing a credible baseline for cyber-physical security analysis.

Through systematic experimentation across multiple attack types, defense configurations, topologies, and network scales, we demonstrated that a full quantum defense stack, comprising QKD, Ping-Pong IDS, and QCA token authentication, which effectively neutralizes coordinated FDI and node-spoofing campaigns, achieving $100\%$ attack block rates and reducing EENS to near-baseline levels. A defense ablation study identified QCA token-based command authentication as the single most impactful quantum component, halving EENS beyond what classical defenses alone could achieve. Overhead analysis confirmed that the quantum stack adds approximately $35$ms to mean control-loop latency, with 95th-percentile tail latency remaining below $75$ms, even though having about $25$ms added latency than the classical defense stack.

There are several open future directions to this work. Firstly, all the quantum primitives are mathematical models rather than state-of-the-art simulators. Integrating hardware-in-the-loop QKD devices or coupling with dedicated quantum network simulators would help validate whether the modeled key rates, latency, and error profiles hold under realistic photonic channel conditions, and can help future users include a comparison between several QKD protocol classes as well. Second, the threat model currently assumes a fixed adversary strategy. Extending it to adaptive adversaries that observe defense responses and escalate accordingly, for instance, switching from FDI to key-exhaustion attacks upon detecting QKD deployment, would provide a more rigorous test of the quantum stack's robustness under strategic escalation. Future works can even explore the use of machine-learning enabled adversary, such as a reinforcement learning model, to adapt more efficiently.  Third, while only classical attack vectors (FDI, spoofing, MITM) are evaluated in this study, quantum-specific attacks such as photon-number splitting, Trojan-horse side-channel exploits, and detector blinding could stress-test the quantum layer itself rather than just the classical control plane it protects. While the base of the simulator is shown to be working well, several of these outline future works can improve both the nature and quality of the quantum-augmented microgrid (QuAM) simulator, thus helping us understand the security needs and possible vulnerabilities in future quantum-augmented microgrid deployments.

\section*{Acknowledgment}
This work is partly sponsored by the National Science Foundation (NSF) awards numbers 2324924 and 2324925.

% References
\bibliography{main} % bibliography data in report.bib
\bibliographystyle{spiebib} % makes bibtex use spiebib.bst

\end{document}